\documentclass[namedreferences]{SolarPhysics}
\usepackage[optionalrh]{spr-sola-addons} 
\usepackage{graphicx}                    
\usepackage{color}                       
\usepackage{url}                         

\newcommand{\etal}{{\it et al.}}

\newcommand{\boldvec}[1]{\mbox{{\boldmath $#1$}}}

\newcommand{\aap}{    {\it Astron. Astrophys.}}

\newcommand{\apj}{    {\it Astrophys. J.}}

\newcommand{\gafd}{   {\it Geophys. Astrophys. Fluid Dyn.}}

\newcommand{\jastp}{  {\it J. Atmos. Solar Terr. Phys.}}
\newcommand{\jgr}{    {\it J. Geophys. Res.}}

\newcommand{\solphys}{{\it Solar Phys.}}

\begin{document}

\begin{article}

\begin{opening}

\title{On the Relationship between Equilibrium Bifurcations and 
Ideal MHD Instabilities for Line-Tied Coronal Loops}

\author{T.~\surname{Neukirch}$^{1}$\sep
Z.~\surname{Romeou}$^{1,2}$}

\runningauthor{Neukirch and Romeou}
\runningtitle{MHD bifurcations and instabilities}

\institute{$^{1}$ School of Mathematics and Statistics,
  University of St. Andrews,
  St. Andrews, KY16 9SS,
  United Kingdom
             email: \url{thomas@mcs.st-and.ac.uk}\\
             $^{2}$ Now at: Ministry of Development,
Department of Metrology, Cannigos Square,
101 81 Athens, Greece, email: \url{zrom@gge.gr}\\
             }   

\date{ }

\begin{abstract}

For axisymmetric models for coronal loops the relationship between 
the bifurcation points of magnetohydrodynamic (MHD) equilibrium sequences and the points of 
linear ideal MHD instability is investigated imposing line-tied boundary 
conditions. 
Using a well-studied example based on the Gold-Hoyle equilibrium, it is demonstrated 
that if the equilibrium sequence is calculated using the Grad-Shafranov equation, the 
instability corresponds to the second bifurcation point and not the first bifurcation 
point because the equilibrium boundary conditions allow for modes which are
excluded from the linear ideal stability analysis. This is shown by calculating the bifurcating 
equilibrium branches and comparing the spatial structure 
of the solutions close to the bifurcation point with
the spatial structure of the unstable mode.
If the equilibrium sequence is calculated using Euler potentials the first bifurcation point of the 
Grad-Shafranov case is not found, and the first bifurcation point of the Euler potential description 
coincides with the ideal instability threshold. An explanation of this results in terms of linear 
bifurcation theory is given and the implications for the use of MHD equilibrium bifurcations to 
explain eruptive phenomena is briefly discussed.

\end{abstract}

\keywords{Magnetohydrodynamics; Instabilities; Corona, Structures; Flares, Relation to Magnetic Field}

\end{opening}

\section{Introduction}

Magnetohydrodynamic instabilities of coronal loops are since a long 
time discussed as one of the main theoretical
explanations for solar flares, in particular compact loop flares ({\it e.g.} \opencite{priest82}). 
Traditionally, investigations
of MHD instabilities of coronal loops model these loops as straight cylindrical flux 
tubes of finite length with
line tied boundary conditions at the `photospheric' ends of the flux tubes 
({\it e.g.} \opencite{raadu72}; \opencite{hood:priest79}, \citeyear{hood:priest81};
\opencite{einaudi:vanhoven83}; \opencite{velli:etal90}). 
Such a set-up allows
for a wide variety of relatively simple equilibrium configurations, hence explaining 
its popularity. 

The stability of equilibrium configurations of the above mentioned type has been studied 
for several decades using
the methods of linear MHD stability analysis 
({\it e.g.} \opencite{raadu72}; \opencite{hood:priest79}, \citeyear{hood:priest81};
\opencite{einaudi:vanhoven83}; \opencite{velli:etal90};
\opencite{debruyne:hood89}, \citeyear{debruyne:hood92}; 
\opencite{mikic:etal90}; \opencite{hood:etal94}; \opencite{vanderlinden:hood98}, \citeyear{vanderlinden:hood99}). In recent years the 
investigations have been extended
into the nonlinear regime using large-scale 
MHD simulations ({\it e.g.} \opencite{longbottom:etal96};
\opencite{baty:heyvaerts96}; \opencite{baty97a},
\citeyear{baty97b}, \citeyear{baty00a}, \citeyear{baty00b}; 
\opencite{lionello:etal98}; \opencite{arber:etal99}; \opencite{gerrard:etal01}; \opencite{browning:vanderlinden03}; \opencite{browning:etal08}; \opencite{hood:etal09}).

In the present contribution we want to investigate the stability of 
line-tied coronal loop models from a
different point of view. The flux tube equilibria used to model coronal 
loops all depend on one or more
parameters representing quantities like the magnetic twist or the plasma beta. 
Many investigations study
how the linear stability of the loops changes as one (or more) of these 
equilibrium parameters vary. 

The systematic variation of one or several parameters of an equilibrium 
defines an equilibrium sequence, and a 
point of linear instability should correspond
to a bifurcation point of the equilibrium sequence and vice versa.
It has to be kept in mind, however, that magnetostatic equilibria
are usually calculated by solving a mathematically reduced set
of equations. It is not at all clear whether there is
really a one-to-one correspondence between
points of
linear instability and bifurcation points, in particular
if line-tied boundary conditions are imposed as in
models of coronal loops.

In the present paper we shall investigate the question
whether the points of linear instability of rotationally
symmetric straight line-tied flux tubes have a one-to-one 
correspondence with the bifurcation points of
equilibrium sequences.

We shall use two different ways of calculating the equilibrium
sequences, namely Grad-Shafranov theory and Euler potentials,
and we shall, for simplicity, investigate only axisymmetric
instabilities and bifurcations. A particularly
well-studied equilibrium class \cite{gold:hoyle60} will be
used to carry out this investigation, mainly
because results of linear stability investigations
for this equilibrium class are readiliy available in the
literature ({\it e.g.} \opencite{hood:priest79}, \citeyear{hood:priest81};
\opencite{mikic:etal90};
\opencite{debruyne:hood92}).

In Section \ref{basic} the basic equilibrium theory and
those parts of the theory of linear MHD stability
needed in this paper are discussed. The following
Section \ref{numerics} presents a brief outline
of the numerical method used to calculate
the equilibrium sequences and to determine
their bifurcation points and bifurcating branches.
The results of these calculations are 
given in Section \ref{results} and discussed
in Section \ref{discussion}. The paper
closes with a summary in Section \ref{summary}.

\section{Basic Theory}

\label{basic}

\subsection{The Gold-Hoyle Equilibrium}
\label{goldhoyle}

We start our investigation from static equilibrium solutions of the
MHD equations, {\it i.e.} solutions of
\begin{eqnarray}
{\bf j}\times{\bf B} - \nabla p & = & {\bf 0} \label{forcebal} \\
\nabla \times {\bf B} & = & \mu_0 {\bf j} \label{ampere}\\
\nabla \cdot {\bf B} &=& 0 \label{divb}.
\end{eqnarray}
We are looking for solutions in cylindrical 
coordinates $r$, $\phi$, $z$, and restrict the spatial domain
to $0 \le z \le L$. The solutions will be considered as
straight flux tube approximations of coronal loops, in the sense
of a large aspect ratio expansion. In this case
$L$ is the loop length and the boundaries $z=0$
and $z=L$ have to be identified with the photospheric
end points of the loop. The centre of the loop is
given by $r=0$.
In the present paper we will only
consider solutions which do not depend on $\phi$,
{\it i.e.} axisymmetric solutions. 

We normalise the magnetic field to the value of $B_z$ 
in the centre of the loop ($r=0$), $B_0$, the coordinates and the
loop length by a typical radial length scale, $b$, and the
pressure by $B_0^2/\mu_0$.
In this normalisation,
the Gold-Hoyle equilibrium \cite{gold:hoyle60} is 
given by the magnetic field components  (see {\it e.g.} \opencite{longbottom:etal96})
\begin{eqnarray}
B_r     & = & 0 ,\\
B_{\phi}& = & \frac{r}{1+r^2} , \label{ghbpr}\\
B_z     & = & \frac{\lambda}{1+r^2}, \label{ghpr}
\end{eqnarray}
and the plasma pressure
\begin{equation}
 p = \frac{1}{2}\frac{1-\lambda^2}{(1+r^2)^2}.
\end{equation}
The parameter $\lambda$ controls both the field line twist
$\Phi$ between $z=0$ and $z=L$,
\begin{equation}
\Phi =  \frac {L}{\lambda}
\end{equation}
and the plasma beta. For $\lambda =1$, the equilibrium
is force-free, {\it i.e.} the current density is parallel to
the magnetic field lines, whereas for $\lambda=0$ the
current density is everywhere perpendicular to the
magnetic field lines. For values of $\lambda$ between
$0$ and $1$ we have a combination of field-aligned
and perpendicular current density. The equilibrium
class is not defined for $\lambda >1$ because the pressure
would become negative in this case.

The Gold-Hoyle equilibrium class depends only on the variable
$r$ and is therefore a one-dimensional MHD equilibrium. 
One-dimensional equilibria of this type can be easily calculated
(see {\it e.g.} \opencite{priest82}, chapter 3.3)
We use
it here as a kind of prototype flux tube equilibrium, because 
linear stability results are readily available for the Gold-Hoyle
equilibrium class \cite{debruyne:hood92}. To obtain
genuinely two-dimensional equilibria depending on $r$ and $z$
we have to resort to one of the more general theories
described in the next sections.

\subsection{Grad-Shafranov Theory}
\label{gradshaf}

To satisfy the solenoidal condition (\ref{divb}),
we write the magnetic field in the form
\begin{equation}
{\bf B} = \frac{1}{r}\nabla A \times {\bf e}_\phi + B_\phi {\bf e}_\phi .
\label{bflux}
\end{equation}
Here the flux function $A$ and the $\phi$-component of the magnetic field
depend only on $r$ and $z$. Taking the scalar product of
Equation (\ref{forcebal}) with {\bf B} and using the
fact that under the condition of
axisymmetry the pressure also depends only on $r$ and $z$,
we find that the pressure is a function of $A$~:
\begin{equation}
p(r,z) = F(A(r,z)).
\end{equation}
An investigation of the $\phi$-component of (\ref{forcebal}) shows
that 
\begin{equation}
b_\phi(r,z) = r B_\phi(r,z,) = G(A(r,z))
\end{equation}
is also a function of $A$ only.  

The force balance equation can then be reduced to
a single partial differential equation for $A$ ({\it e.g.} \opencite{bateman78})~:
\begin{equation}
-r\nabla \cdot \left(\frac{1}{r^2}\nabla A\right) = r \frac{d p}{d A} + \frac{1}{r} b_\phi \frac{d b_\phi}{d A}.
\label{gseq}
\end{equation}

The dependence of the pressure and the $\phi$-component of the magnetic
field on the flux function $A$ have to be specified for a solution of this equation. For the
Gold-Hoyle solution the flux function $A$ is given by
\begin{equation}
A_{\rm GH}(r,z) = \frac{\lambda}{2}\ln(1+r^2).
\label{ghfluxf}
\end{equation} 
Using equation (\ref{ghfluxf}) to determine $r$ as a function of $A$, 
and substituting this expression into Equations   (\ref{ghbpr}) and (\ref{ghpr})
we find that
\begin{eqnarray}
p       &=& \frac{1}{2}(1-\lambda^2) \exp\left(-\frac{4}{\lambda} A\right), \label{ghpa}\\
b_\phi  &=& 1-\exp\left(-\frac{2}{\lambda} A\right)                    .    \label{ghbpa}
\end{eqnarray}

For each value of $\lambda$, the Gold-Hoyle equilibrium is therefore a 
solution of the Grad-Shafranov equation
\begin{eqnarray}
-\frac{\partial}{\partial r}\left(\frac{1}{r}\frac{\partial A}{\partial r}\right)
-\frac{1}{r}\frac{\partial^2 A}{\partial z^2 } &= &
-2 r  \frac{1-\lambda^2}{\lambda}  \exp\left(-\frac{4}{\lambda} A\right) \nonumber \\
& & \mbox{\hspace{-1cm}}
+
\frac{2}{\lambda\, r} \left[1-\exp\left(-\frac{2}{\lambda} A\right)\right]
\exp\left(-\frac{2}{\lambda} A\right),
\label{gsgh}
\end{eqnarray}
if the boundary conditions at $z=0$ and $z=L$ are given by $A=\lambda/2\ln(1+r^2)$ as well.
As the Gold-Hoyle solutions depend on the parameter $\lambda$
they define a solution branch of the Grad-Shafranov Equation (\ref{gsgh}).
As this equation is nonlinear it is to be expected that it can
also have other solution branches for the same boundary conditions.
Points where the Gold-Hoyle solution branch and any other solution branches meet
are called bifurcation points. Standard bifurcation theory ({\it e.g.} \opencite{iooss:joseph80})
tells us that at bifurcation points the stability of the solution branches
can change. We will discuss this possibility in more detail in Section
\ref{stability}.

\subsection{Euler Potentials}
\label{euler}

The Grad-Shafranov theory is very useful for symmetric plasma systems, but a Grad-Shafranov
type equation can only be derived for translational, rotational and helical symmetry 
(\opencite{solovev67}; \opencite{edenstrasser80a}, \citeyear{edenstrasser80b}). Without such a symmetry,
we have to use a different way to calculate MHD equilibria. The approach coming
closest to the use of a flux function for symmetric systems is to use
Euler potentials (sometimes also called Clebsch coordinates) to describe
the magnetic field. The Euler potential approach has the advantage that it
can be used to describe three
dimensional magnetic fields without symmetry, although there are
some restrictions concerning the existence of Euler potentials for
given magnetic fields
({\it e.g.} \opencite{hesse88}; \opencite{rosner:etal89}). For
the flux tube like equilibria considered in the present paper
these constraints do not apply.

Another reason for using Euler potentials even for symmetric cases
is that certain types of constraints are a lot easier
to impose with Euler potentials then with a Grad-Shafranov description.
A typical example from solar physics is the quasi-static
shearing of magnetic arcades. In this case the footpoint displacement of
the fieldlines is the physical parameter which is determined by
the boundary conditions. In this case the use of
Euler potentials is very useful 
({\it e.g.} \opencite{barnes:sturrock72}; \opencite{zwingmann87}; 
\opencite{platt:neukirch94}; \opencite{antiochos:etal99}).

With the Euler potentials $\alpha$ and $\beta$ a general magnetic field can be written
as
\begin{equation}
{\bf B}=\nabla\alpha \times \nabla \beta.
\label{beulergen}
\end{equation}
Any vector field of this form is automatically solenoidal. In the present paper we restrict
our analysis to axisymmetric fields. In this case the Euler potential $\alpha$ is a
function of $r$ and $z$ only, the Euler potential $\beta$ is
chosen as $\beta(r,\phi,z) = \tilde{\beta}(r,z) + \phi$ and
Equation (\ref{beulergen}) reduces to
\begin{equation}
{\bf B}=\frac{1}{r}\nabla\alpha \times {\bf e}_\phi +\nabla\alpha \times \nabla \tilde{\beta}.
\label{beuleraxi}
\end{equation}
By comparison with Equation (\ref{bflux}), we see that now the Euler potential $\alpha$ 
corresponds to the flux function $A$, whereas $B_\phi$ has been replaced by the
$
\nabla\alpha \times \nabla \tilde{\beta}
$.
Substitution of Equation (\ref{beuleraxi}) into Equation (\ref{forcebal}) and
(\ref{ampere}) gives the two equations for $\alpha$ and $\tilde{\beta}$
({\it e.g.} \opencite{zwingmann87}; \opencite{platt:neukirch94})~:
\begin{eqnarray}
\nabla \tilde{\beta}\cdot\nabla\times(\nabla\alpha\times\nabla\tilde{\beta})-
\nabla\cdot\left(\frac{1}{r^2}\nabla \alpha \right)
& = & \frac{d p}{d\alpha}, \label{eulera}\\
\nabla\alpha\cdot\nabla\times(\nabla\tilde \beta\times\nabla\alpha) &= &0.
\label{eulerb}
\end{eqnarray}
For the Gold-Hoyle solution $\alpha_{\rm GH}$ is identical with the flux function
$A_{\rm GH}$ given in Equation (\ref{ghfluxf}). The pressure function
$p(\alpha)$ has the same form as $p(A)$ in Equation (\ref{ghpa}), 
only with $\alpha$ replacing $A$.
We can use $B_\phi$ to work out that for the Gold-Hoyle solution
\begin{equation}
\tilde {\beta}_{\rm GH} = -\frac{1}{\lambda} z .
\label{betatgh}
\end{equation}
The function $\tilde{\beta}$ represents the fieldline twist for the
Gold-Hoyle solution since
\begin{equation}
\tilde{\beta}_{\rm GH}(r,0) - \tilde{\beta}_{\rm GH}(r,L) = \frac{L}{\lambda} = \Phi
\end{equation}

We impose boundary conditions for both Euler potentials. The boundary
conditions for $\alpha$ are the same as for $A$ in the Grad-Shafranov case.
For $\tilde{\beta}$ we use Equation (\ref{betatgh}) on the boundaries. This
fixes the footpoint displacement of fieldlines crossing the
boundaries $z=0$ and $z=L$. In the same way as in the Grad-Shafranov case
the Gold-Hoyle solutions are a solution branch for the  
Equations (\ref{eulera}) and (\ref{eulerb}). The same statements about
bifurcations and stability apply as in the Grad-Shafranov case.

\subsection{Linear Stability}
\label{stability}

The theory of linear MHD stability is a vast area and we only summarise some
results which are important for the following discussion.
Defining the Lagrangian displacement $\boldvec{\xi}$ the linearized
ideal MHD equations can be written as
\begin{equation}
\rho_0 \frac{\partial^2 \boldvec{\xi}}{\partial t^2 } = {\bf F}(\boldvec{\xi})
\label{linearmhd}
\end{equation}
where
\begin{equation}
{\bf F}(\boldvec{\xi}) = \frac{1}{\mu_0}[(\nabla\times{\bf B}_1)\times{\bf B}_0]
                        +\frac{1}{\mu_0}[(\nabla\times{\bf B}_0)\times{\bf B}_1]
                        +\nabla(\boldvec{\xi}\cdot\nabla p_0 + \gamma p_0\nabla\cdot\boldvec{\xi}).
\end{equation}
The components of the magnetic field perturbation ${\bf B}_1$ are given by
\begin{equation}
{\bf B}_1 = \nabla\times(\boldvec{\xi}\times{\bf B}_0).
\label{linearb1}
\end{equation}
For the problem we are discussing in the
present paper Equation (\ref{linearmhd}) has to be solved 
on a tube-like domain with line-tying boundary conditions at
$z=0$ and $z=L$, with ${\bf B}_0 $ and $p_0$ given by the
Gold-Hoyle solution. The line-tying condition 
corresponds to
\begin{equation}
{\bf \boldvec{\xi}} = {\bf 0}
\end{equation}
on the boundaries. Assuming an exponential time-dependence for the
perturbation ${\boldvec{\xi}}$, one obtains a self-adjoined eigenvalue problem.
Instabilities occur when one of the eigenvalues of the equation
changes sign. The corresponding perturbations $\boldvec{\xi}$ can be
classified according to their different spatial structure. In general
we speak of different modes when refering to the spatial structure
of the instabilities. For each mode it is usually sufficient to investigate
the largest eigenvalue corresponding to this mode, as it is this
eigenvalue which determines whether a mode is stable or unstable.

It can be shown 
that the Gold-Hoyle
solution is
always stable for $\lambda=1$, {\it i.e.} in the force-free case.
When decreasing $\lambda$ the value of $\lambda$
 where the Gold-Hoyle solutions become unstable to the
different possible modes under line-tying boundary
conditions depends on the length
of the flux tube $L$. A thorough investigation of this problem
has been carried
out by\inlinecite{debruyne:hood92} and we will make use of their
results in the later parts of this paper. Since we investigate
only axisymmetric equilibria and bifurcations we will also restrict
our attention to the axisymmetric modes (sometimes called ``sausage modes'').

\section{Numerical Method}
\label{numerics}

The numerical calculations have been carried out with a code based 
on a 
continuation method ({\it e.g.} \opencite{allgower:georg90}).
The code used here is based on a method proposed by\inlinecite{keller77}
and has been successfully applied to a variety of
problems in plasma physics, solar physics, magnetospheric physics and
astrophysics 
({\it e.g.}
\opencite{zwingmann83}, \citeyear{zwingmann87};
\opencite{neukirch93a}, \citeyear{neukirch93b};
\opencite{neukirch:hesse93}; \opencite{platt:neukirch94}; \opencite{schroer:etal94}; \opencite{becker:etal96}, \citeyear{becker:etal01}; \opencite{romeou:neukirch99}, \citeyear{romeou:neukirch01}, \citeyear{romeou:neukirch02}, \citeyear{romeou:neukirch02b}; \opencite{kiessling:neukirch03}).
The method has the advantage that it can calculate sequences
of equilibria depending on an external parameter (like
$\lambda$ for the Gold-Hoyle solutions), and detect bifurcation points.
It is also possible to calculate bifurcating equilibrium sequences.
The code uses a finite element discretization allowing for a flexible
grid structure. Further details can be found in\inlinecite{neukirch93a}
and\inlinecite{neukirch93b}.

We have solved both the Grad-Shafranov Equation (\ref{gsgh}) and the
Euler potential Equations (\ref{eulera}) and (\ref{eulerb}) on a
numerical domain extending from $r=0$ to $r=8$ and from $z=0$ to
$z=L$, where $L$ is varied between $3$ and $8$. The radial extent of
the domain is chosen along the same lines as done by\inlinecite{longbottom:etal96} in their MHD simulations of the
sausage instability.

For the Grad-Shafranov equilibrium sequences we have used 
\begin{equation}
A_b=A_{\rm GH}
\end{equation}
as boundary condition on all boundaries. In the Euler potential case the boundary conditions
are given by
\begin{eqnarray}
\alpha_b&=&A_{\rm GH}, \\
\tilde{\beta}_b &=& \tilde{\beta}_{\rm GH} .
\end{eqnarray}
Note that in both cases both the differential equation and the boundary conditions 
depend on the parameter $\lambda$. Due to the boundary conditions
the Gold-Hoyle solutions are one solution branch of the equations. This
can be used to check the accuracy of the numerical code and to adjust the
resolution. In all runs presented in Section \ref{results} we have used
a numerical grid with 1800 triangular finite elements corresponding to 
a resolution of 61 by 61 grid points in each spatial direction. The grid
is equidistant in the $z$-direction but non-equidistant in the $r$-direction
with a higher resolution towards the axis of the tube.

\section{Results}
\label{results}

For both the Grad-Shafranov and the Euler potential case we have carried
out a numerical investigation of the bifurcation properties of
the Gold-Hoyle solution branch using the numerical method described in
Section \ref{numerics}. For a series of values of the 
loop length $L$, we have first calculated the Gold-Hoyle
branch with our code, starting with the force-free solution ($\lambda=1$)
and then following the branch for decreasing $\lambda$ into
the non-force-free regime. Although we know the Gold-Hoyle branch analytically
this procedure allows us to check the accuracy of our numerical
calculations and to use the capability of the code to
detect bifurcation points. At such points other solution branches cross
the Gold-Hoyle branch, and we expect that those points correspond
to the instability threshold of the $m=0$-instability under
line-tying conditions as, for example, calculated by\inlinecite{debruyne:hood92}. For the detected bifurcation
points we have then also calculated the bifurcating branches
for a range of $\lambda$ values. This is important to check
whether the spatial structure of the bifurcating solution branch
coincides with the predictions made by linear stability theory
on the basis of the structure of the unstable mode.

An important point to emphasize here is that because we calculate
the Gold-Hoyle branch in the direction of decreasing $\lambda$,
we will also number the bifurcation points in this direction, {\it i.e.}
when we speak of first and second bifurcation the $\lambda$ value
of the first bifurcation will be bigger than the $\lambda$ value of the
second bifurcation. Although this is opposite to the
terminology normally used in bifurcation theory, we have 
decided to keep the parametrization used by\inlinecite{debruyne:hood92} 
to make a comparison
with their results easier.

\subsection{The Grad-Shafranov Case}

We have carried out calculations of the Gold-Hoyle branch for loop
lenghts $L=$ $3.0$, $4.0$, $5.0$, $6.0$, $7.0$, $8.0$, $9.0$ and $10.0$.
In all cases, we have calculated the Gold-Hoyle branch 
until we had found at least two bifurcation points.
The $\lambda$ values of the two first bifurcation points
found by code for the different $L$ values are
listed in Table \ref{GSbp}.
\begin{table}[t]
\caption{The first and second bifurcation points for the Grad-Shafranov case.}
\begin{tabular}{ccccccccc}
\hline
$L$&3.0&4.0&5.0&6.0&7.0&8.0&9.0&10.0\\
\hline
$\lambda_1$& 0.598&0.629&0.652&0.663&0.675&0.684& 0.684&0.690\\
$\lambda_2$&0.472 & 0.528&0.566 &0.593 &0.612 &0.627 & 0.639&0.648 \\
\hline
\end{tabular}
\label{GSbp}
\end{table}
\begin{figure}
\begin{center}

\includegraphics[width=0.95\textwidth]{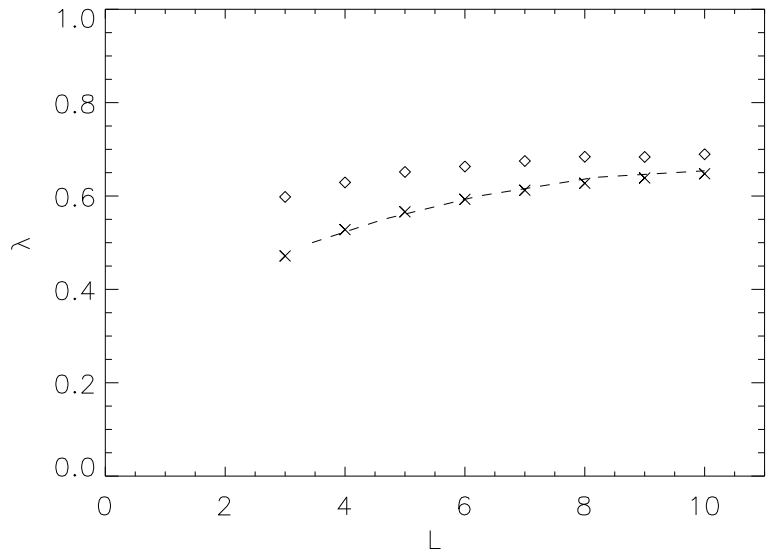}
\end{center}
\caption{The dependence of the $\lambda$ values
of the first ($\diamond$) and the second ($\times$) bifurcation point
on the loop length $L$
for the Grad-Shafranov case. The dashed line is the
instability threshold of the $m=0$ mode derived by 
\protect\inlinecite{debruyne:hood92} using
linear MHD stability theory under line-tying
conditions. Solutions with $\lambda$ values below the dashed
line are unstable with respect to the sausage mode.
It is obvious that
the second
and not the first bifurcation point for a given loop length 
corresponds to the $m= 0$ instability.
}
\label{gsstabl}
\end{figure}
A graphical representation of these values is shown in 
Figure \ref{gsstabl}. In this figure we plot the 
values for the first and second bifurcation points
in the $L$-$\lambda$-plane. Also shown in Figure
\ref{gsstabl} is the stability threshold 
for the $m=0$-instability (dashed line) derived by
\inlinecite{debruyne:hood92} using
linear MHD theory.
The figure clearly shows that the linear stability threshold
corresponds to the second bifurcation along the
Gold-Hoyle branch. This raises the question 
what the first bifurcation point corresponds to.

\begin{figure}
\begin{center}
\includegraphics[width=0.8\textwidth]{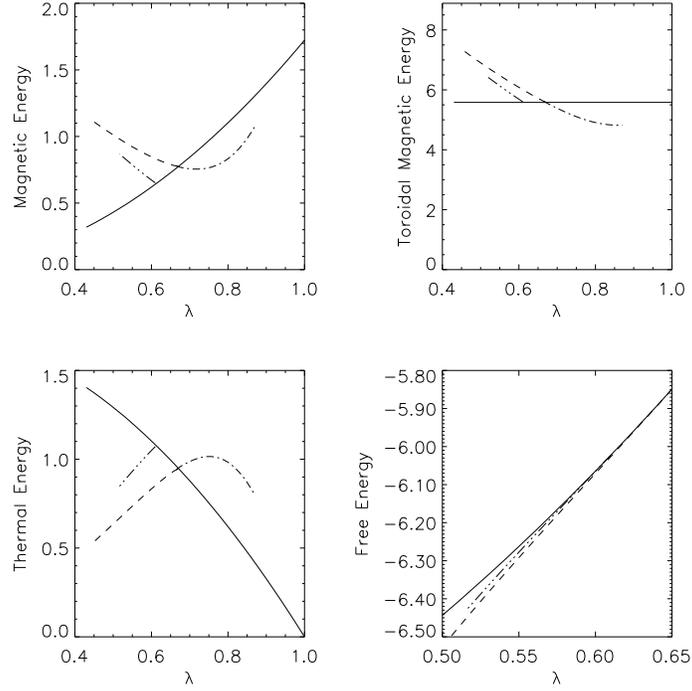}
\end{center}
\caption{Bifurcation diagrams of the Grad-Shafranov case 
for $L=7.0$. Shown are the poloidal magnetic energy (upper left),
the toroidal magnetic energy (upper right), the thermal energy (lower left), and the free energy
as defined by \protect\inlinecite{grad64}
(lower right). For definitions of these quantities see the main text. The
free energy is only shown for values of $\lambda$ close to the second bifurcation point to
make the difference between the branches more obvious.}
\label{bifdiaggs7}
\end{figure}

To answer this question, we have calculated the bifurcating branches for
the first and second bifurcation points for loop lengths of
$L=3.0$, $5.0$ and $7.0$. The 
structure of the bifurcation diagrams is very similar for all
three cases and we therefore only show the case $L=7.0$
(Figure \ref{bifdiaggs7}).
The four quantities shown in Figure \ref{bifdiaggs7} are the polodial magnetic energy
\begin{equation}
W_{\rm p} = \int\frac{1}{2} \left(\frac{1}{r}\nabla A \right)^2 d V,
\end{equation}
the toroidal magnetic energy
\begin{equation}
W_{\rm t} = \int\frac{1}{2} \left(\frac{1}{r} b_\phi \right)^2 d V,
\end{equation}
the thermal energy
\begin{equation}
W_{\rm th} = \int p\, d V,
\end{equation}
and the free energy defined by\inlinecite{grad64}
\begin{equation}
W_{\rm f} = W_{\rm p} -(W_{\rm t}+W_{\rm th}) .
\end{equation}
At the first bifurcation point another solution branch crosses the 
Gold-Hoyle branch. The bifurcating branch exists for values of $\lambda$
both smaller and larger than the bifurcation point $\lambda$. At the bifurcation point
the poloidal and toroidal magnetic energies of the bifurcating branch go
from values smaller than the energies of the Gold-Hoyle branch to values
larger than the Gold-Hoyle branch in the direction of decreasing $\lambda$. 
The thermal energy of the bifurcating branch 
is higher than that of the Gold-Hoyle branch for $\lambda$ larger than the
bifurcation $\lambda$ and smaller than the Gold-Hoyle branch beyond the
bifurcation point.

The second bifurcating solution branch only exists for values of $\lambda$
which are smaller than the bifurcation $\lambda$. This is to be expected
on the basis of standard bifurcation theory ({\it e.g.} \opencite{iooss:joseph80}),
taking the spatial structure of the solutions along the branch into account (see below). 
The mathematical argumentation is given in the Appendix. The poloidal and
toroidal magnetic energies along this branch are larger than those of the 
Gold-Hoyle branch, whereas the thermal energy is smaller than the thermal energy
of the Gold-Hoyle branch for the same value of $\lambda$.

Also shown in Figure \ref{bifdiaggs7} is a plot of the free energy for values
of $\lambda$ close to the second bifurcation point to enhance the difference between the branches. 
We can see that in this range the Gold-Hoyle branch has
the biggest free energy. The first bifurcating branch has lower free energy than
the second branch but both branches have a lower free energy than the
Gold-Hoyle branch. This shows that a transition to the bifurcating branches at fixed $\lambda$ is 
indeed energetically favourable for the system, as the system will always try to
settle into a state of lower free energy.

\begin{figure}[t]
\begin{center}
\includegraphics[width=0.45\textwidth]{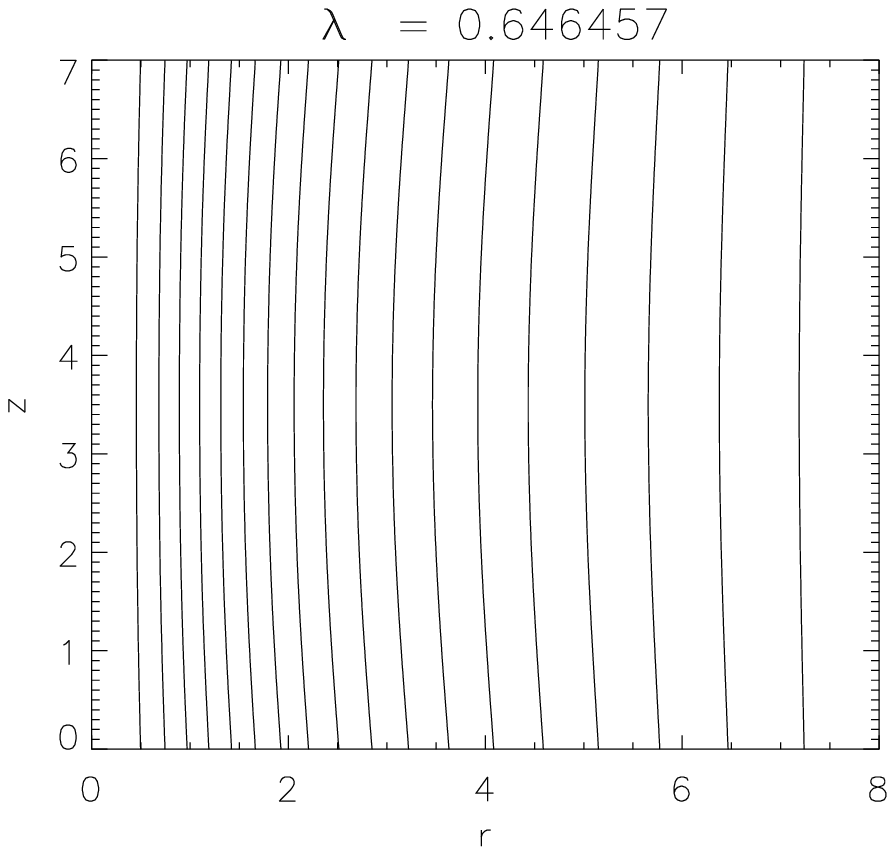} \hfil 
\includegraphics[width=0.45\textwidth]{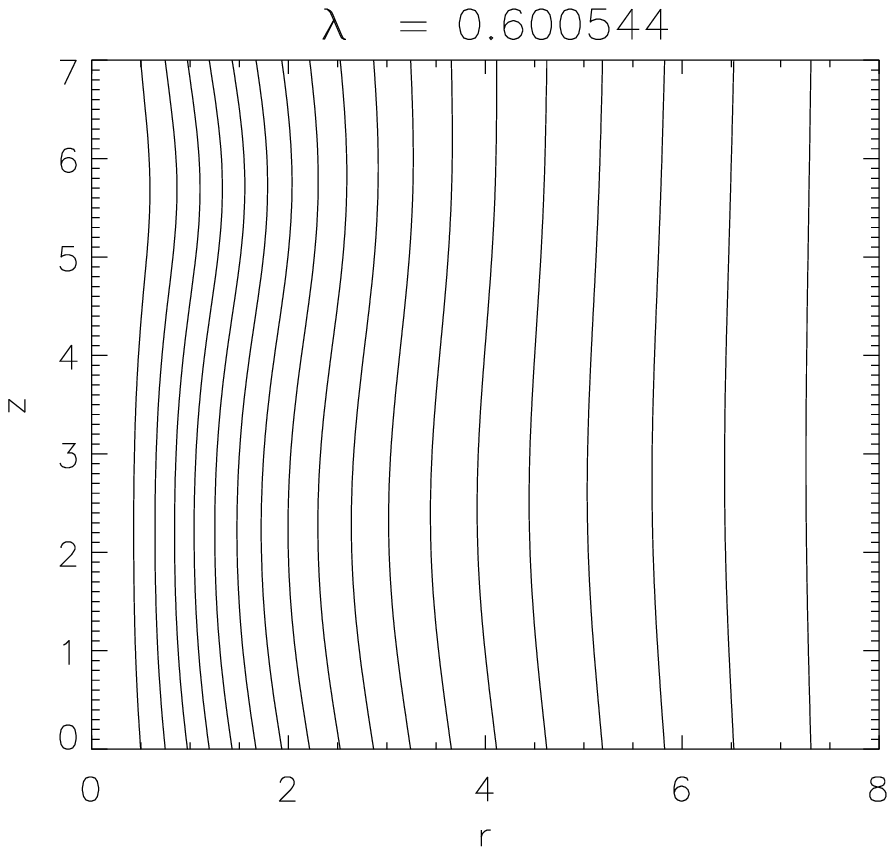}
\end{center}
\caption{Left: Solution for $\lambda= 0.646457 $ on first bifurcating branch. Right:
Solution for $\lambda= 0.600544 $ on second bifurcating branch. Whereas the solutions
on the first bifurcating branch show a $\sin(\pi z/L)$-dependence superimposed on the
Gold-Hoyle solution, the 
solutions on the second bifurcating branch have a $\sin(2\pi z/L)$-dependence.}
\label{bifsolgs5}
\end{figure}
The spatial structure of the solutions on the bifurcating branches
is shown in Figure \ref{bifsolgs5}.
The obvious difference between the solutions on the two branches is their
dependence on $z$. Whereas the solutions on the first bifurcating branch show
a $\sin(\pi z/L)$-dependence superimposed on the
Gold-Hoyle solution, the 
solutions on the second bifurcating branch have a $\sin(2\pi z/L)$-dependence.
Both functions are consistent with the boundary condition $A=A_{\rm GH}$ at
$z=0$ and $z=L$. We will discuss the implications of this finding in the light
of linear stability theory in Section \ref{discussion}.

\subsection{The Euler Potential Case}

In the Euler potential case we have carried calculations of the Gold-Hoyle branch
for the same values $L$ as in the Grad-Shafranov case. The calculations were run
for about the same $\lambda$ range as for the Grad-Shafranov equation, but only one bifurcation
point was detected in this range. The $\lambda$ values of the bifurcation
point for all loop lengths is given in Table \ref{EPbp}. A comparison with
Table \ref{GSbp} shows that the first bifurcation point in the Euler
potential case corresponds to the second bifurcation point
of the Grad-Shafranov case.
\begin{table}[t]
\caption{The first bifurcation point for the Euler potential case.}
\begin{tabular}{ccccccccc}
\hline
$L$&3.0&4.0&5.0&6.0&7.0&8.0&9.0&10.0\\
\hline
$\lambda_1$& 0.472 & 0.528&0.566 &0.593 &0.613 &0.628 & 0.639&0.648 \\
\hline
\end{tabular}
\label{EPbp}
\end{table}
\begin{figure}
\begin{center}
\includegraphics[width=0.95\textwidth]{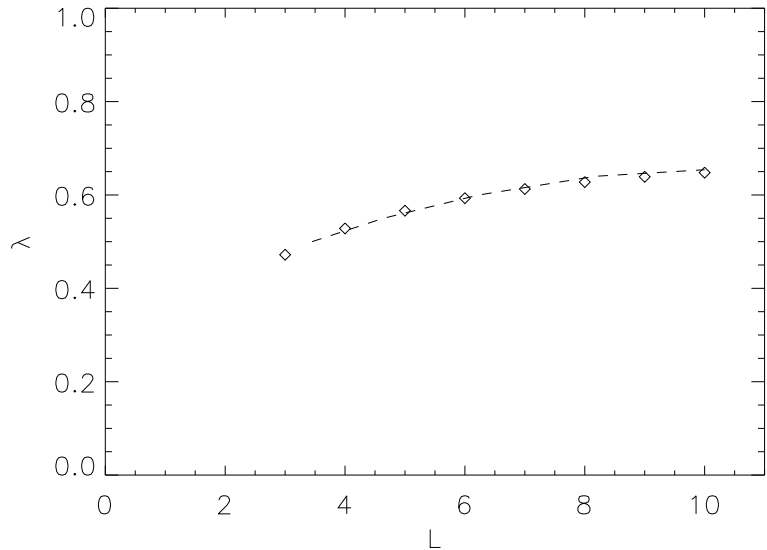}
\end{center}
\caption{The dependence of the $\lambda$ values on the loop length $L$ for
the Euler potential case. The Euler potential case is different from
 the  Grad-Shafranov case because in this case the first ($\diamond$)  
bifurcation point corresponds to the point where the
$m= 0$-mode becomes unstable.}
\label{epstabl}
\end{figure}

This is also obvious if we plot
the $\lambda$ values of the bifurcation point for different $L$ in the
$L$-$\lambda$-plane to compare with the results of
\inlinecite{debruyne:hood92} (see Figure \ref{epstabl}).
It can clearly be seen that
in the Euler potential case it is obviously the first bifurcation which coincides
with the stability threshold of linear MHD. This difference between the Grad-Shafranov
case and the Euler potential case is surprising and we will discuss the reasons for
this in Section \ref{discussion}.

\begin{figure}
\begin{center}
\includegraphics[width=0.8\textwidth]{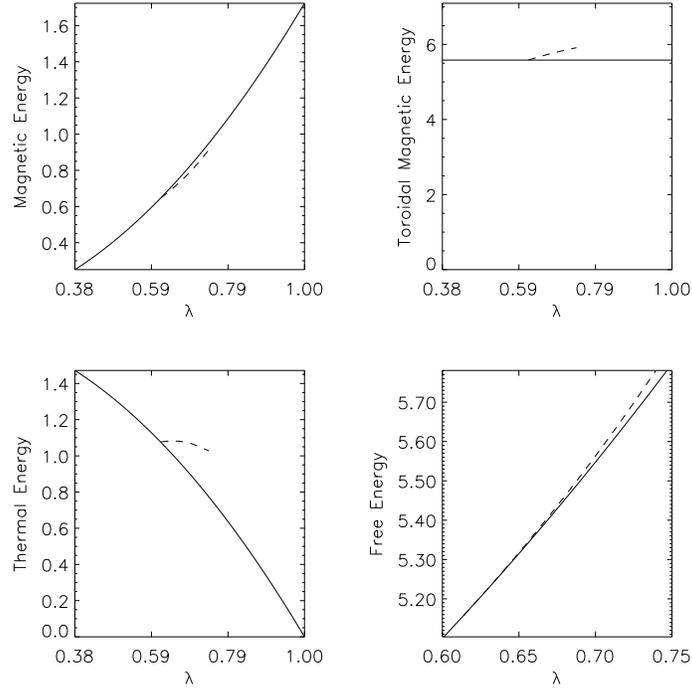}
\end{center}
\caption{Bifurcation diagrams of the Euler potential case 
for $L=7.0$. Shown are the poloidal magnetic energy (upper left),
the toroidal magnetic energy (upper right), the thermal energy (lower left) and free energy
(lower right). The major differences to Figure \ref{bifdiaggs7} are that the bifurcation point shown in this diagram corresponds to the second bifurcation point of the Grad-Shafranov case, and that the bifurcating solution sequence branches off in the direction of increasing $\lambda$. This opposite to the Grad-Shafranov case.}
\label{bifdiagep7}
\end{figure}
Another major difference between the Grad-Shafranov case and the Euler potential case is that in the Euler potential case the new solution sequence branches off towards increasing values of $\lambda$, whereas for the Grad-Shafranov case the bifurcating sequence branching off towards decreasing 
$\lambda$ values. This has implication for the stability of the bifurcating branch.

For the Euler potential case we define the poloidal magnetic energy as
\begin{equation}
W_{\rm p} = \int\frac{1}{2} \left(\frac{1}{r}\nabla \alpha \right)^2 d V,
\end{equation}
the toroidal magnetic energy as
\begin{equation}
W_{\rm t} = \int\frac{1}{2} \left(\nabla \alpha \times \nabla \tilde{\beta} \right)^2 d V,
\end{equation}
the thermal energy as
\begin{equation}
W_{\rm th} = \int p\, d V,
\end{equation}
and the free energy as
\begin{equation}
W_{\rm f} = W_{\rm p} + W_{\rm t}-W_{\rm th}.
\end{equation}
One should note that the contribution of the toroidal magnetic energy to the free energy is positive for the Euler potential case, whereas it is negative in the Grad-Shafranov case. The reason for this are the different constraints on the system in the two cases (see \opencite{grad64}).
The poloidal magnetic energy of the bifurcating branch is slightly lower than that of the Gold-Hoyle solution, whereas the toroidal magnetic energy is higher. The thermal energy is also higher than the thermal energy of the Gold-Hoyle solution and the free energy is slightly larger than that of the Gold-Hoyle solution. 

\begin{figure}
\begin{center}
\includegraphics[width=0.8\textwidth]{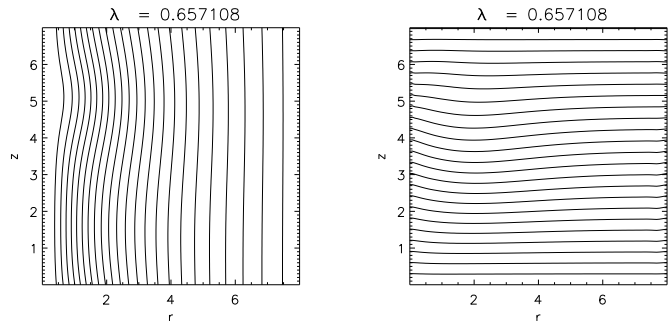}
\end{center}
\caption{Contour plot of the Euler potentials $\alpha$ (left) and $\tilde{\beta}$ (right) on the bifurcating branch for $\lambda= 0.657$. The spatial structure of $\alpha$ is similar to the spatial structure of $A$ in the Grad-Shafranov case.}
\label{L7bifep}
\end{figure}
The spatial structure of $\alpha$ on the bifurcating branch is similar to the spatial structure of $A$ in the Grad-Shafranov case (see Figure \ref{L7bifep}; obviously there is no analogue for $\tilde{\beta}$ in the Grad-Shafranov case). In the linear regime  $\alpha$ has a  $\sin(2\pi z/L)$ $z$-dependence, 
whereas $\tilde{\beta}$ has a $1-\cos(2\pi z/L)$ structure due to Equation (\ref{eulerb}).

\section{Discussion}
\label{discussion}

The results presented in Section \ref{results} raise the following questions. 
\begin{itemize}

\item Why does the
$m=0$-instability correspond to the second and not to the first bifurcation point
in the Grad Shafranove case ? 

\item Why is the Euler potential case different from the Grad-Shafranov case ?

\end{itemize}

To answer these questions we first analyse the connection between the line-tying condition
in linear ideal MHD stability and the boundary conditions in
the Grad-Shafranov and Euler potential cases. 
Close to the bifurcation points, we can represent
the solutions on the bifurcating branches for the Grad-Shafranov case by
\begin{equation}
A_{\rm bif}(r,z,\lambda) = A_{\rm GH}(r,\lambda) + \epsilon A_1(r,z) + \ldots \;,
\label{abifgs}
\end{equation}
where $\epsilon \ll 1$. Note that this expansion differs slightly
from the expansion used in the Appendix.
To first order in $\epsilon$ the function $A_1$ has to satisfy the equation
\begin{equation}
-\nabla \cdot \frac{1}{r^2} \nabla A_1 = \left( r \frac{d^2 p}{d A^2}\Big|_{A_{\rm GH}} +
                                         \frac{1}{2r} \frac{d^2 b_\phi^2}{d A^2}\Big|_{A_{\rm GH}}
                                         \right) A_1,
\label{A1eq}
\end{equation}
where $p$ and $b_\phi$ are given by Equations (\ref{ghpa}) and (\ref{ghbpa}).
Since the boundary conditions for $A_{\rm bif}$ are already satisfied by $A_{\rm GH}$, the
function $A_1$ must vanish on all boundaries. Since all coefficients of 
Equation (\ref{A1eq}) depend only on $r$, it is easy to see that the solutions of Equation (\ref{A1eq})
must be have the form
\begin{equation}
A_1(r,z) = F_n(r) \sin( n \pi z/L), \quad n=1,2,3,\ldots \;,
\label{A1form}
\end{equation}
with $F_n(r)$ a radial function. 

Close to the bifurcation point, a linear stability analysis of the fundamental Gold-Hoyle branch
would give Lagrangian perturbations $\boldvec{\xi}$ which are related to $A_1$ through Equation (\ref{linearb1}).
Since the deviation of the poloidal field from the Gold-Hoyle branch along the bifurcating
branch is given by $\nabla \times (A_1 \nabla \phi)$, one can easily see that
\begin{equation}
 A_1 \nabla\phi = -(\boldvec{\xi} \cdot \nabla A_{\rm GH}) \nabla \phi 
\end{equation}
so that
\begin{equation}
A_1 = - \xi_r \frac{\partial A_{\rm GH}}{\partial r}
\label{A1lintheory}
\end{equation}
since $A_{\rm GH}$ depends only on $r$.
Equation (\ref{A1lintheory}) shows that the boundary condition $A_1=0$ only implies
$\xi_r=0$, but not necessarily $\boldvec{\xi}={\bf 0}$. We therefore surmise that the first bifurcation
in the Grad-Shafranov case corresponds to a linear Lagrangian displacement with 
non-vanishing $\xi_\phi$ and/or $\xi_z$. The solutions on the first
branch would satisfy the boundary condition
$A_1 =0$, but not $\boldvec{\xi}={\bf 0}$.
The first bifurcating branch therefore corresponds to solutions which do not
satisfy the line-tying boundary conditions, and, for example, can only be reached
from the Gold-Hoyle branch if flow through the boundary is allowed ($\xi_z \ne 0$).

The second bifurcating branch satisfies
both $A_1=0$ and $\boldvec{\xi}={\bf 0}$. This is corroborated by the fact that the
$z$-dependence of the $r$-component of the
Lagrangian perturbation for the sausage mode is given by
$\sin(2\pi z/L)$
({\it e.g.} 
\opencite{longbottom:etal96}). This matches exactly the $z$-dependence
of $A_1$ on the second bifurcating branch. Therefore the
bifurcation points and the linear instability thresholds coincide.

In the Euler potential case, we have boundary conditions for both $\alpha$
and $\tilde{\beta}$, thus constraining the system more than in the Grad-Shafranov case.
One can derive the connection between the Lagrangian perturbation $\boldvec{\xi}$ and the linear perturbations $\alpha_1$ and $\tilde{\beta}_1$ of the Euler potential from the expression for the linear perturbation of the magnetic field,
\begin{equation}
\mathbf{B}_1 = \nabla \times (\boldvec{\xi}\times \mathbf{B}_0) = \nabla \alpha_1 \times \nabla \phi +
\nabla \alpha_1 \times \nabla \beta_0 + \nabla \alpha_0 \times \nabla \tilde{\beta}_1.
\label{linearB}
\end{equation}
With a bit of algebra one can show ({\it e.g.} \opencite{zwingmann87}) that
\begin{equation}
\alpha_1 = - \boldvec{\xi}\cdot\nabla\alpha_0, \quad \tilde{\beta}_1 = -\boldvec{\xi}\cdot\nabla (\phi + \tilde{\beta}_0),
\label{linearEuler}
\end{equation}
which for axisymmetric flux tube equilibria like the Gold-Hoyle equilibrium discussed in this paper leads to
\begin{equation}
\alpha_1 = - \xi_r \frac{\partial \alpha_0}{\partial r}, \quad 
\tilde{\beta}_1 = -\frac{1}{r}\xi_\phi - \xi_z \frac{\partial \tilde{\beta}_0}{\partial z}.
\label{linearEuler2}
\end{equation}
As is to be expected the expression connecting $\alpha_1$ and $\xi_r$ is the same as for $A_1$ 
and $\xi_r$ in the Grad-Shafranov case, and $\alpha_1 =0$ on the boundaries ensures that $\xi_r=0$ on the boundaries. The boundary condition $\beta_1 =0$ imposes an additional constraint, which links $\xi_\phi$ and $\xi_z$ on the boundaries, ensuring that $\boldvec{\xi}_\perp$ vanishes on the boundaries. This is consistent with the line-tying boundary conditions imposed by\inlinecite{debruyne:hood92} and explains why the bifurcation points coincide with the linear stability threshold in the Euler potential case.

We suspect, but cannot prove, that the different structure of the bifurcation diagrams present in Figures
\ref{bifdiaggs7} and \ref{bifdiagep7} is also due to the different constraints imposed upon the system by using different descriptions for the magnetic field. The different structure of the bifurcation diagrams may have implication for the stability of the bifurcating equilibrium branch.  Usually, when moving along a stable equilibrium sequence and crossing a bifurcation point so that the equilibrium sequence is unstable beyond the bifurcation point, the bifurcating branch is linearly stable close to the bifurcation point if it bifurcates in the forward direction and linearly unstable if it bifurcates in the backward direction (see {\it e.g.} \opencite{iooss:joseph80}). In the present case this would imply that in the Grad-Shafranov case the second bifurcating branch is linearly stable, whereas this branch is unstable in the Euler potential case. This is also supported by the fact that the second bifurcation branch has a lower free energy than the Gold-Hoyle branch for the Grad-Shafranov case, whereas it has a higher free-energy in the Euler potential case. It has to be remarked, however, that this is a conjecture as we have no rigorous proof.

\section{Summary and Conclusion}
\label{summary}

We have investigated the relationship between MHD bifurcation and linear stability for a class of axisymmetric straight flux tubes under line-tying boundary conditions. For simplicity we only considered rotationally symmetric perturbations, allowing only for sausage modes. We have used two different ways of calculating the equilibrium sequences including bifurcating branches - one approach uses the Grad-Shafranov equation, the other approach uses Euler potentials. It turns out that only the Euler potential case shows a one-to-one correspondence between the first bifurcation point and the linear instability threshold for the sausage mode. The Grad-Shafranov case shows an additional bifurcation which does not correspond to the instability threshold under line-tying boundary conditions. This difference can be explained by the different constraints imposed on the bifurcating equilibrium branches in the Grad-Shafranov and the Euler potential cases. 

Furthermore, even though the second bifurcation point of the Grad-Shafranov case coincides with the first bifurcation point of the Euler potential case and the linear instability threshold, the structure of the bifurcation diagrams differ considerably between the Grad-Shafranov and the Euler potential case. 
The reason for this is not yet clear, but is probably also due to the difference in boundary conditions.
In any case this difference has implications for the stability of the bifurcating equilibrium branches (see {\it e.g.} \opencite{iooss:joseph80}) and is therefore important to decide whether the system is able to find a new equilibrium (in the present case a new axisymmetric equilibrium) if one would consider an imaginary process driving the flux tube across the instability threshold.

The present investigation is a preparation for studying equilibrium sequences of magnetic  flux tubes and other solar magnetic structures together with their bifurcations in three dimensions. Preliminary steps have already been made (see {\it e.g.} \opencite{romeou:neukirch02}) and more detailed investigations are planned for the future.

\begin{acks}

The authors thank Alan Hood for useful discussions. 
T. Neukirch acknowledges support by STFC and by the European Commission through the SOLAIRE Network (MTRN-CT-2006-035484). 
Z. Romeou gratefully acknowledges financial support 
provided through the European Community's Training and Mobility of Researchers Programme
by
a Marie-Curie Fellowship and through the European Community's
Human Potential Programme under contract HPRN-CT-2000-00153, PLATON.
The
authors also acknowledge partial support by the British Council ARC Programme.

\end{acks}

\begin{appendix}


Whereas the first bifurcating branch in the Grad-Shafranov case exists for values of $\lambda$
which are both bigger and smaller than the $\lambda$ at the bifurcation point,
the second bifurcating branch exists only for $\lambda$ smaller than the
bifurcation $\lambda$. This fact can be explained by using standard bifurcation
theory to calculate the structure of the bifurcating branches close to the
bifurcation points. The argument is actually independent of the form of the
functions $p(A)$ and $b_\phi(A)$. The qualitative structure of the bifurcation
diagram will thus be the same even if $p(A)$ and $b_\phi(A)$ are changed as long
as the fundamental branch consists of solutions which depend only on the radial
coordinate $r$.

We start by writing the Grad-Shafranov equation in the form
\begin{equation}
G(A,\lambda,r)=-r\nabla\cdot\left(\frac{1}{r^2}\nabla A\right) - N(A,\lambda,r) = 0,
\label{gsgeneral}
\end{equation}
where the function 
$N(A,r,\lambda)$ summarizes the nonlinear part of the Grad-Shafra\-nov equation given
by $p(A,\lambda)$ and $b_\phi(A,\lambda)$. For the present paper
$p$ and $b_\phi$ are given by Equations (\ref{ghpa}) and (\ref{ghbpa}). For the following
argument, however, the exact form of $N(A,r,\lambda)$ is irrelevant, as long as it is
analytic in $A$ and $\lambda$ at the bifurcation points we want to investigate. We will not
give here any details of the mathematical background which can be found for example in
\inlinecite{hesse:ks86} and\inlinecite{hesse:kiessling87}. These papers treat slighly
different bifurcation problems, but we will be using the same technique.

Let $\lambda^*$ be the value of $\lambda$ at either of the bifurcation points
and let $A_0=A_0(\lambda^*)$ be the solution of Equation (\ref{gsgeneral})
at the bifurcation point. To calculate the bifurcating
branch we expand $\lambda$ and $A$ as
\begin{eqnarray}
\lambda &=& \sum_{k=0}^\infty \epsilon^k \lambda_k ,\label{lambdaexp} \\
A &=& \sum_{k=0}^\infty \epsilon^k A_k        ,      \label{Aexp}
\end{eqnarray}
where $\lambda_0=\lambda^*$ and $A_0$ as above. Since $G$ is analytic
in both $A$ and $\lambda$ for $\lambda > 0$ we can expand Equation (\ref{gsgeneral})
in a power series in $\epsilon$:
\begin{equation}
0 = \sum_{k=0}^\infty \frac{1}{k !}\frac{d^k }{d\epsilon^k} 
G(A(\epsilon),\lambda(\epsilon),r )\Big|_{\epsilon=0} \epsilon^k .
\end{equation}
As each power of $\epsilon$ must satisfy this equation independently we obtain
\begin{equation}
\frac{d^k }{d\epsilon^k} 
G(A(\epsilon),\lambda(\epsilon),r )\Big|_{\epsilon=0} = 0, \qquad k=0,1,2,\ldots\; .
\label{Gexp}
\end{equation}
Obviously, the lowest order equation
\begin{equation}
G(A_0(\lambda^*),\lambda^*,r) = 0
\end{equation}
is just the Grad-Shafranov equation at the bifurcation point and therefore
trivially satisfied.

For the discussion of the higher order equations we first have to look at the
boundary conditions the $A_k$ have to satisfy. The boundary condtion $A_b$ for $A(r,z,\lambda)$
is given by the fundamental branch solution $A_b(r,z,\lambda)=A_0(r,z,\lambda)$
(the Gold-Hoyle solution in the present paper). 
Therefore we can extend $A_b$ into the domain. Using the expansion (\ref{lambdaexp}) in
$A_b(r,z,\lambda(\epsilon))$ we can see that the boundary condition each of the $A_k$ in
Equation (\ref{Aexp}) has to satisfy is given by
\begin{equation}
A_b^{(k)} = \frac{1}{k !}\frac{d^k}{d \epsilon^k} A_0(r,z,\lambda(\epsilon)) \Big|_{\epsilon=0}.
\end{equation}
Note that $A_b^{(k)}$ satifies the same Equation (\ref{Gexp}) as $A_k$.

For $O(\epsilon)$ we get from Equation (\ref{Gexp}) 
\begin{equation}
G_A(A_0(\lambda^*),\lambda^*, r) A_1 + G_\lambda(A_0(\lambda^*),\lambda^*, r) \lambda_1 =0,
\end{equation}
with
\begin{eqnarray}
G_A A_1 &=& -r\nabla\cdot\left(\frac{1}{r^2}\nabla A_1\right) -
\frac{\partial N}{\partial A}(A_0,\lambda^*,r) A_1 ,\\
G_\lambda &=&\frac{\partial N}{\partial \lambda}(A_0,\lambda^*,r).
\end{eqnarray}
As mentioned above $A_b^{(1)}$ satisfies the same equation as $A_1$ and therefore the
function
\begin{equation}
A_1^\prime = A_1 -A_b^{(1)}
\end{equation}
satisfies $G_A A_1^\prime =0$ or explicitely
\begin{equation}
 -r\nabla\cdot\left(\frac{1}{r^2}\nabla A_1^\prime\right) -
\frac{\partial N}{\partial A}(A_0,\lambda^*,r) A_1^\prime =0
\label{firstorder}
\end{equation}
with $A_1^\prime = 0$ on the boundaries.
Since all coefficients of Equation (\ref{firstorder}) depend only on $r$ its
solution can be obtained by separation of variables with the
general
form of $A_1^\prime$ being
\begin{equation}
A_1^\prime(r,z) = F_n(r) \sin(n\pi z/L), \qquad n= 1,2,3,\ldots \;.
\label{a1modes}
\end{equation}
The exact form of $F_n(r)$ is of no importance for the following argument.

If we want to calculate $\lambda_1$, we have to go to the next order ($O(\epsilon^2)$)
of the
expansion, giving
\begin{eqnarray}
& &-r\nabla\cdot\left(\frac{1}{r^2}\nabla A_2\right)-\frac{\partial N}{\partial A} A_2
-\frac{1}{2}\frac{\partial^2 N}{\partial A^2} A_1^2 
- \frac{\partial^2 N}{\partial A \partial \lambda} A_1 \lambda_1 \nonumber \\
& &\mbox{ \hspace{5cm}}-\frac{1}{2}\frac{\partial^2 N}{\partial \lambda^2} \lambda_1^2
 - \frac{\partial N}{\partial \lambda} \lambda_2 =0
\end{eqnarray}
where all derivatives of $N(A,\lambda,r)$ are evaluated at the bifurcation point ($\epsilon=0$).
Similarly to $A_b^{(1)}$ at $O(\epsilon)$, $A_b^{(2)}$ satisfies the same equation as $A_2$.
We define 
\begin{equation}
A_2^\prime=A_2-A_b^{(2)}
\end{equation}
which obeys the equation
\begin{equation}
G_A A_2^\prime = \frac{1}{2}\frac{\partial^2 N}{\partial A^2} 
({A_1^\prime}^2 + 2 A_b^{(1)} A_1^\prime) +
\lambda_1\frac{\partial^2 N}{\partial A \partial \lambda} A_1^\prime .
\label{secondorder}
\end{equation}
By Fredholm's alternative the right hand side of Equation
(\ref{secondorder}) has to be orthogonal to $A_1^\prime$, {\it i.e.}
\begin{equation}
\int_0^L\int_0^{r_{max}}\left(
\frac{1}{2}\frac{\partial^2 N}{\partial A^2} 
({A_1^\prime}^2 + 2 A_b^{(1)} A_1^\prime) +
\lambda_1\frac{\partial^2 N}{\partial A \partial \lambda} A_1^\prime \right)
A_1^\prime rdrdz = 0 .
\label{fredha}
\end{equation}
To proceed we assume in agreement with the Gold-Hoyle solution 
that the function $A_b^{(1)}$ has the form
\begin{equation}
A_b^{(1)} = \lambda_1 f_b(r)
\end{equation}
where $f_b(r)$ is left unspecified here.
Equation (\ref{fredha}) can then be used to calculate $\lambda_1$
in the form
\begin{eqnarray}
\lefteqn{
\lambda_1 \int_0^L\int_0^{r_{max}}\left(
\frac{\partial^2 N}{\partial A^2} f_b(r) +
\frac{\partial^2 N}{\partial A \partial \lambda}\right)
{A_1^\prime}^2 r dr dz  =   }   \nonumber \\
& & \mbox{\hspace{4cm}}-\int_0^L\int_0^{r_{max}}
\frac{1}{2}\frac{\partial^2 N}{\partial A^2} 
{A_1^\prime}^3 r dr dz .
\label{l1equat}
\end{eqnarray}
The double integral on the right hand side of Equation (\ref{l1equat})
can be split into two separate integrations over $r$ and $z$, since
the integrand depends on $z$ only through
$A_1^\prime$. As $A_1^\prime$ has the
form (\ref{a1modes}), the integral over $z$ is given by
\begin{equation}
\int_0^L \sin^3(n\pi z/L) dz = -\frac{L}{n\pi}(\cos n\pi -1) +\frac{L}{3n\pi}(\cos^3 n\pi -1) ,
\end{equation}
which vanishes for all even $n$. As the integral on the left hand side
of Equation (\ref{l1equat}) is nonzero, this implies that for even
$n$ (and in particular for $n=2$) $\lambda_1$ vanishes. The bifurcation
at bifurcation points with modes having even $n$ is therefore
quadratic.

This explains the structure of the bifurcation diagram in Figure \ref{bifdiaggs7}, because
the first bifurcation obviously corresponds to the $A_1^\prime$ for
$n=1$, whereas the second bifurcation corresponds to $n=2$. Therefore the structure
of the first branch close to the bifurcation point is given by
\begin{eqnarray}
\lambda & = &\lambda^* + \epsilon \lambda_1 + \ldots, \label{lambdabif1} \\
A       & = & A_0 (r,z,\lambda^*) + \epsilon A_1(r,z,\lambda^*) + \ldots.\label{Abif1}
\end{eqnarray}
The slope of the bifurcating branch at the bifurcation point is determined by
$\lambda_1 \ne 0$ in this case and it is obvious that the bifurcating branch exists
for both $\lambda > \lambda^*$ ($\epsilon \lambda_1 > 0$) and $\lambda < \lambda^*$
($\epsilon \lambda_1 < 0$).

Close to the second bifurcation point we have
\begin{eqnarray}
\lambda & = &\lambda^* + \frac{1}{2}\epsilon^2 \lambda_2 + \ldots, \label{lambdabif2}\\
A       & = & A_0 (r,z,\lambda^*) + \epsilon A_1(r,z,\lambda^*) + \ldots,  \label{Abif2}
\end{eqnarray}
because here $\lambda_1$ vanishes. As the correction to $\lambda^*$ depends quadratically
on $\epsilon$, positive and negative $\epsilon$  give the same value of $\lambda$. This implies
that the second bifurcating branch actually consists of {\em two} branches, one for positive and
one for negative $\epsilon$. Since a change of sign of $\epsilon$ in 
Equation (\ref{Abif2}) corresponds
to a simple mirroring of the $\sin(2\pi z/L)$ function at the point $z=L/2$, the two branches have
exactly the same energies. We remark that since $\lambda_1=0$ in this case $A_1 = A_1^\prime$ as
the boundary contribution to $A_1^\prime$ vanishes. The numerical calculations corroborate these
results as the same second bifurcation branch is found by the code starting both with
negative and positive $\epsilon$. The only difference between the calculations is the mirroring
of the $z$-dependence of $A$ along the bifurcating branch.

\end{appendix}

\end{article}

\begin{thebibliography}{}

\bibitem[\protect\citeauthoryear{Allgower and Georg}{1990}]{allgower:georg90}
Allgower, E. L., Georg, K.: 1990, {\it Numerical Continuation Methods},
Springer, Berlin, p.~7.

\bibitem[\protect\citeauthoryear{Antiochos \etal}{1999}]{antiochos:etal99}
Antiochos, S. K., DeVore, C. R., Klimchuk, J. A.: 1999,
{\apj}  {\bf 510}, 485.

\bibitem[\protect\citeauthoryear{Arber \etal}{1999}]{arber:etal99}
Arber, T. D., Longbottom, A. W., Van der Linden, R. A. M.: 1999,
{\apj}  {\bf 517}, 990.

\bibitem[\protect\citeauthoryear{Barnes and Sturrock}{1972}]{barnes:sturrock72}
Barnes, C. W., Sturrock, P. A.: 1972,
{\apj} {\bf 174}, 659.

\bibitem[\protect\citeauthoryear{Bateman}{1978}]{bateman78}
Bateman, G.: 1978, {\it MHD Instabilities}, MIT Press, Cambridge MA, p.~68.

\bibitem[\protect\citeauthoryear{Baty}{1997a}]{baty97a}
Baty, H.: 1997a,
{\aap} {\bf 318}, 621.

\bibitem[\protect\citeauthoryear{Baty}{1997b}]{baty97b}
Baty, H.: 1997b
{\solphys} {\bf 172}, 249.

\bibitem[\protect\citeauthoryear{Baty}{2000a}]{baty00a}
Baty, H.: 2000a
{\aap} {\bf 353}, 1074.

\bibitem[\protect\citeauthoryear{Baty}{2000b}]{baty00b}
Baty, H.: 2000b
{\aap} {\bf 360}, 345.

\bibitem[\protect\citeauthoryear{Baty and Heyvaerts}{1996}]{baty:heyvaerts96}
Baty, H., Heyvaerts, J.: 1996
{\aap} {\bf 308}, 935.

\bibitem[\protect\citeauthoryear{Becker \etal}{1996}]{becker:etal96}
Becker, U., Neukirch, T., Birk, G. T.: 1996, {\it Phys. Plasmas} {\bf 3}, 1452.

\bibitem[\protect\citeauthoryear{Becker \etal}{2001}]{becker:etal01}
Becker, U., Neukirch, T., Schindler, K.: 2001, {\jgr} {\bf 106},
3811.

\bibitem[\protect\citeauthoryear{Browning \etal}{2008}]{browning:etal08}
Browning, P.K., Gerrard, C., Hood, A.W., Kevis, R., and van der Linden, R.A.M.: 2008, 
{\aap} {\bf 485}, 837. 

\bibitem[\protect\citeauthoryear{Browning and Van der 
Linden}{2003}]{browning:vanderlinden03}
Browning, P.K.~and Van der Linden, R.A.M.: 2003, {\aap} {\bf 400}, 355. 

\bibitem[\protect\citeauthoryear{De Bruyne and Hood}{1989}]{debruyne:hood89}
De Bruyne, P., Hood, A. W.: 1989, {\solphys} {\bf 119}, 87.

\bibitem[\protect\citeauthoryear{De Bruyne and Hood}{1992}]{debruyne:hood92}
De Bruyne, P., Hood, A. W.: 1992, {\solphys} {\bf 142}, 87.

\bibitem[\protect\citeauthoryear{Edenstrasser}{1980a}]{edenstrasser80a}
Edenstrasser, J. W.: 1980a, {\it J. Plasma Phys.} {\bf 24}, 299. 

\bibitem[\protect\citeauthoryear{Edenstrasser}{1980b}]{edenstrasser80b}
Edenstrasser, J. W.: 1980b, {\it J. Plasma Phys.} {\bf 24}, 515.

\bibitem[\protect\citeauthoryear{Einaudi and Van Hoven}{1983}]{einaudi:vanhoven83}
Einaudi, G., Van Hoven, G,: 1983,
{\solphys} {\bf 88}, 163.

\bibitem[\protect\citeauthoryear{Gerrard \etal}{2001}]{gerrard:etal01}
Gerrard, C. L., Arber, T. D., Hood, A. W., Van der Linden, R. A. M.:
2001, {\aap} {\bf 373}, 1089.

\bibitem[\protect\citeauthoryear{Grad}{1964}]{grad64}
Grad, H.: 1964, {\it Phys. Fluids} {\bf 7}, 1283.

\bibitem[\protect\citeauthoryear{Gold and Hoyle}{1960}]{gold:hoyle60}
Gold, T., Hoyle, F.: 1960, {\apj} {\bf 120 }, 89.

\bibitem[\protect\citeauthoryear{Hesse}{1988}]{hesse88}
Hesse, M.: 1988, {\it Untersuchungen zur magnetischen
Rekonnektion in dreidimensionalen Systemen }, Ph.D. Thesis, Ruhr-Universit\"at Bochum.

\bibitem[\protect\citeauthoryear{Hesse and Kiessling}{1987}]{hesse:kiessling87}
Hesse, M., Kiessling, M.: 1987, {\it Phys. Fluids} {\bf 30}, 2720.

\bibitem[\protect\citeauthoryear{Hesse and Schindler}{1986}]{hesse:ks86}
Hesse, M., Schindler, K.: 1986, {\it Phys. Fluids} {\bf 29}, 2484.

\bibitem[\protect\citeauthoryear{Hood \etal}{2009}]{hood:etal09}
Hood, A. W., Browning, P.K., Van der Linden, R. A. M.: 2009 {\aap}, submitted.

\bibitem[\protect\citeauthoryear{Hood and Priest}{1979}]{hood:priest79}
Hood, A. W., Priest, E. R.: 1979, 
{\solphys} {\bf 64}, 303.

\bibitem[\protect\citeauthoryear{Hood and Priest}{1981}]{hood:priest81}
Hood, A. W., Priest, E. R.: 1981,
{\gafd} {\bf 17}, 297.

\bibitem[\protect\citeauthoryear{Hood \etal}{1994}]{hood:etal94}
Hood, A. W., De Bruyne, P., Van der Linden, R. A. M., Goossens, M.: 1994, 
{\solphys} {\bf 150}, 99.

\bibitem[\protect\citeauthoryear{Iooss and Joseph}{1980}]{iooss:joseph80}
Iooss, G., Joseph, D. D.: 1980, {\it Elementary Stability and Bifurcation Theory},
Springer, New York, p.~42.

\bibitem[\protect\citeauthoryear{Keller}{1977}]{keller77}
Keller, H. B.: 1977, in P. H. Rabinowitz (ed.), {\it Applications of Bifurcation Theory},  
Academic Press, New York, p. 359.

\bibitem[\protect\citeauthoryear{Kiessling and Neukirch}{2003}]{kiessling:neukirch03}
Kiessling, M.K.-H.~and Neukirch, T.: 2003, {\it Proc.~Natl.~Acad.~Sci.} {\bf 100}, 1510.

\bibitem[\protect\citeauthoryear{Lionello \etal}{1998}]{lionello:etal98}
Lionello, R., Velli, M., Einaudi, G., Miki\'{c}, Z.: 1998,
{\apj} {\bf 494}, 840.

\bibitem[\protect\citeauthoryear{Longbottom \etal}{1996}]{longbottom:etal96}
Longbottom, A.W., Hood, A.W., Rickard, G.: 1996, {\it 
Plasma Phys. Control. Fusion} {\bf 38}, 193.

\bibitem[\protect\citeauthoryear{Miki\'{c} \etal}{1990}]{mikic:etal90}
Miki\'{c}, Z., Schnack, D. D., Van Hoven, G.: 1990,
{\apj} {\bf 361}, 690.

\bibitem[\protect\citeauthoryear{Neukirch}{1993a}]{neukirch93a}
Neukirch, T.: 1993a, {\jgr} {\bf 98}, 3753.

\bibitem[\protect\citeauthoryear{Neukirch}{1993b}]{neukirch93b}
Neukirch, T.: 1993b, {\aap} {\bf 274}, 319.

\bibitem[\protect\citeauthoryear{Neukirch and Hesse}{1993}]{neukirch:hesse93}
Neukirch, T.~and Hesse, M.: 1993, {\apj} {\bf 411}, 840. 

\bibitem[\protect\citeauthoryear{Priest}{1982}]{priest82}
Priest, E. R.: 1982, {\it Solar Magnetohydrodynamics}, D. Reidel, Dordrecht.

\bibitem[\protect\citeauthoryear{Platt and Neukirch}{1994}]{platt:neukirch94}
Platt, U., Neukirch, T.: 1994, {\solphys} {\bf 153}, 287.

\bibitem[\protect\citeauthoryear{Raadu}{1972}]{raadu72}
Raadu, M. A.: 1972, {\solphys} {\bf 22}, 425.


\bibitem[\protect\citeauthoryear{Romeou and Neukirch}{1999}]{romeou:neukirch99}
Romeou, Z.~and Neukirch, T.: 1999, 
In: Wilson, A. (ed.), {\it Magnetic Fields and Solar Processes, ESA-SP 448}, 
871. 

\bibitem[\protect\citeauthoryear{Romeou and Neukirch}{2001}]{romeou:neukirch01}
Romeou, Z.~and  Neukirch, T.: 2001, 
In: Hanslmeier, A., Messerotti, M., Veronig, A. (eds.),
{\it The Dynamic Sun}, Kluwer Academic Publishers, Dordrecht, 
303. 

\bibitem[\protect\citeauthoryear{Romeou and Neukirch}{2002a}]{romeou:neukirch02}
Romeou, Z.~and 
Neukirch, T.: 2002a, {\jastp} {\bf 64}, 639. 

\bibitem[\protect\citeauthoryear{Romeou and Neukirch}{2002b}]{romeou:neukirch02b}
Romeou, Z.~and 
Neukirch, T.: 2002b, 
In: Sawaya-Lacoste, H. (ed.),
{\it Proc. SOLMAG: Magnetic Coupling 
of the Solar Atmosphere Euroconference, ESA-SP 505}, 549. 

\bibitem[\protect\citeauthoryear{Rosner \etal}{1989}]{rosner:etal89}
Rosner, R., Low, B. C., Tsinganos, K., Berger, M. A.: 1989,
{\gafd} {\bf 48}, 251.

\bibitem[\protect\citeauthoryear{Schindler at al.}{1983}]{schindler:etal83}
Schindler, K.,  Birn, J., Janicke, L.: {\solphys} {\bf 87}, 103.

\bibitem[\protect\citeauthoryear{Schr\"oer \etal}{1994}]{schroer:etal94}
Schr\"oer, A., Neukirch, T., Kiessling, M. K.-H., Hesse, M., Schindler, K.: 1994,
{\it Phys. Plasmas} {\bf 1}, 213.


\bibitem[\protect\citeauthoryear{Solovev}{1967}]{solovev67}
Solovev, L. S.: 1967, in M. A. Leontovich (ed.), {\it Reviews of Plasma Physics} {\bf 3}, 
Consultans Bureau, New York, p. 277.

\bibitem[\protect\citeauthoryear{van der Linden and 
Hood}{1998}]{vanderlinden:hood98}
van der Linden, R.A.M.~and Hood, A.W.: 1998, {\aap} {\bf 339}, 887. 

\bibitem[\protect\citeauthoryear{van der Linden and 
Hood}{1999}]{vanderlinden:hood99}
van der Linden, R.A.M.~and Hood, A.W.: 1999, {\aap} {\bf 346}, 303. 

\bibitem[\protect\citeauthoryear{Velli \etal}{1990}]{velli:etal90}
Velli, M., Hood, A. W., Einaudi, G.: 1990,
{\apj} {\bf 350}, 428.

\bibitem[\protect\citeauthoryear{Zwingmann}{1983}]{zwingmann83}
Zwingmann, W.: 1983,
{\jgr} {\bf 88}, 9101.

\bibitem[\protect\citeauthoryear{Zwingmann}{1987}]{zwingmann87}
Zwingmann, W.: 1987,
{\solphys} {\bf 111}, 309.

\end{thebibliography}
\end{document}